\documentclass[12pt]{article}
\usepackage[active]{srcltx}
\usepackage{graphicx}

\newcommand{\be}{\begin{equation}}
\newcommand{\dd}{\displaystyle}
\newcommand{\ee}{\end{equation}}
\newcommand{\bea}{\begin{eqnarray}}
\newcommand{\eea}{\end{eqnarray}}

\begin{document}
\begin{center}
{\Large\bf\boldmath {Jet analysis by Deterministic Annealing}} \rm
\vskip1pc {\large L. Angelini$^{a,b,c}$, P. De Felice$^{a,b}$, M.
Maggi$^{c}$, G. Nardulli$^{a,b,c}$, L.Nitti$^{b,c,d}$,
M.Pellicoro$^{a,b,c}$ and S. Stramaglia$^{a,b,c}$}\\ \vspace{5mm}
{\it{
$^{a}$Dipartimento Interateneo di Fisica, Bari, Italy \\
 $^{b}$TIRES, Center of Innovative Technologies for Image Detections and Processing, Bari,
Italy\\ $^{c}$I.N.F.N., Sezione di Bari, Italy \\$^{d}$DETO,
Universit\`a di Bari, Italy}}
 \end{center}
 %%% ----------------------------------------------------------------------

 \begin{abstract}
We perform a comparison of two jet clusterization algorithms. The
first one is the standard Durham algorithm and the second one is a
global optimization scheme, {\it Deterministic Annealing}, often
used in clusterization problems, and adapted to the problem of jet
identification in particle production by high energy collisions;
in particular we study hadronic jets in $WW$ production by high
energy $e^+e^-$ scattering. Our results
 are as follows. First, we find that the two procedures give basically
the same output as far as the particle clusterization is
concerned. Second, we find that the increase of CPU time with the
particle multiplicity is much faster for the Durham jet clustering
algorithm  in comparison with Deterministic Annealing. Since this
result follows from  the higher computational complexity of the
Durham scheme, it should not depend on the particular process
studied here and  might be significant for jet physics at LHC as
well.\vskip.5cm
 \par\noindent PACS Numbers: 13.87.-a, 13.38.Be, 05.10.-a, 45.10.Db
\end{abstract}
\section{Introduction}
The problem of clustering consists of optimal grouping of observed signal samples (set of
features). In many circumstances one seeks the partition of the given set of features  which
minimizes a prescribed cost function. This function should embody the {\it a priori} knowledge
on the geometrical aspects of the problem. By this way clustering is thus transformed in an
optimization task. The main applications of clustering are in pattern recognition and signal
compression. Here we address a different problem, i.e. the task of partitioning particles
produced in high energy $e^+e^-$ annihilation into  a certain number of cone regions, i.e.
small solid angles. This is the well known problem of jet clustering in high energy physics. It
arises from the need to relate energy and momentum of the cluster (jet) to the four-momentum of
the underlying and unobservable parton. There are at the moment a few well established jet
clustering algorithms, to be reviewed in section 2. While tuned to the particular problem at
hand they were not studied from the point of view of the theory of clustering algorithms and
are not generally considered as minimization problems of prescribed cost functions. We feel
this is a matter to be further investigated and therefore in this letter we wish to raise a few
questions: Can the existing jet algorithms result from some variational principle? Can the
general theory of clustering algorithms be used to reduce the computational time of jet
algorithms? In this letter we  answer these questions by considering a specific algorithm, the
Deterministic Annealing (DA) \cite{DA,DA2}, to be reviewed in section \ref{sec3}; using an
analogy with statistical physics, DA relates the problem of clustering  to that of finding the
global minimum of a thermodynamical potential. In order to compare this method to the
traditional ones we will study
 a specific issue, i.e. the production of jets in $WW$
 production in high energy $e^+e^-$ colliders. This application will be discussed
 in section \ref{sec4}.
Our results are contained in section \ref{sec5}; they show  that the jet clustering
algorithms and Deterministic Annealing give similar performances, but the computational
complexity of DA is considerably lower;  the use of methods based on variational
principles, such as DA, would be therefore certainly of interest in the analysis of
hadronic final states at the future accelerators
 such as the Large Hadron Collider at CERN, where
the multiplicities might be very high and the computational
complexity extremely demanding
 for ordinary jet clustering algorithms.

\section{Jet clustering algorithms\label{sec2}}
The most common jet clustering algorithms used in studies of $e^+e^-$ annihilation are the JADE
\cite{JADE}, Durham \cite{Durham} and Cambridge \cite{Cambridge} algorithms\footnote{For a
review of these and other jet algorithms see \cite{rassegne2}. For a review of the Montecarlo
generators and their connections with the jet algorithms see \cite{rassegne}. }. The prototype
of these clustering methods and the oldest one is the JADE algorithm. One considers all the
possible pairs $(i,\,j)$ of particles in the final state, with energies $E_i$, $E_j$, and
angular separation $\theta_{ij}$ and computes the jet resolution variable \be
y_{ij}=y^J_{ij}\equiv \frac{2E_iE_j(1-\cos\theta_{ij})}{E^2_{vis}}\ , \label{jadeyij} \ee where
$E_{vis}$ is the visible energy, i.e. the sum of energies for all particles observed in the
final state. This {\it test variable} is then compared to a given threshold parameter $y_{cut}$
and the pair is recombined into a new pseudo-particle $k$ of four-momentum \be p_k=p_i+p_j
\label{Escheme}\ee ($E$ scheme) provided that $y_{ij}\,\leq \,y_{cut}$. The algorithm is then
reiterated to the new set of (pseudo)particles and it stops when, for all pairs, $y_{ij}\,\geq
\,y_{cut}$. The number of pseudoparticles at the end of the algorithm counts the number of
jets, which is therefore fixed by $y_{cut}$. The theoretical advantage of this recombination
scheme lies in the absence of collinear and infrared singularities, as the regions of phase
space where these divergences could be generated are automatically excluded.

A drawback of the definition (\ref{jadeyij}) is that also particles at very different angles
can be recombined in one pseudoparticle. This is due to the fact that
$2E_iE_j(1-\cos\theta_{ij})\approx M^2_{ij}$, where $M^2_{ij}$ is
 the invariant mass of the two particles. On the theoretical side this implies the presence
of large order corrections that cannot be resummed; on the experimental side the trouble is
that {\it ghost} jets may appear i.e. jets along directions where no particles are present. To
cure this problem the Durham algorithm was introduced, which is based on the following
definition of the test variable:
 \be
y_{ij}=y^D_{ij}\equiv\frac{2\,min\{E_i^2,E_j^2\}(1-\cos\theta_{ij})}{E^2_{vis}}\
.\label{durhamyij}\ee
 Clearly the resolution criterion $y^D_{ij}>y_{cut}$ becomes, for small angles,
$k^2_{Ti}>E^2_{vis}\, y_{cut}$, where $k_{Ti}$ is the transverse momentum of the $i-$th
particle to the direction of the $j-$th one. In this way the algorithm tries to minimize the
transverse momentum and not the invariant mass. On the other hand the recombination scheme is
still given by (\ref{Escheme}).

The Durham algorithm presents a problem at very small values of $y_{cut}$. As a matter of fact,
when one tries to resolve the final state to get a larger number of jets, particles that are
almost collinear can be recombined, thus producing unwanted {\it junk} jets. This feature is
solved by the Cambridge algorithm by introducing a third step in the process of formation of
the clusters. Before considering the  jet resolution and the recombination steps, one
introduces an ordering variable $v_{ij}=2(1-\cos\theta_{ij})$. Once the pair $(k,\ell)$ with
the minimum value of $v_{ij}$ is found, the resolution variable $y_{k\ell}$ (still given by
(\ref{durhamyij})) is computed and, after a comparison with $y_{cut}$, the recombination scheme
(\ref{Escheme}) is applied\footnote{The Cambridge algorithm implements the so called {\it soft
freezing}, i.e. if the particles are not recombined the softer particle is removed and
considered as a resolved jet.}.

In this paper we are not interested in the analysis of small $y_{cut}$ as we consider the
production of a $W$ pair in $e^+e^-$ scattering, i.e. a fixed (four) number of jets. In this
context the Durham and the Cambridge algorithms produce analogous results. Therefore in the
sequel we will mainly refer to the Durham algorithm.  For a more detailed comparison among jet
algorithms one can however see \cite{comparison}.
\section{Deterministic annealing\label{sec3}}
Deterministic annealing algorithms \cite{DA} (for reviews see \cite{DA2}) take their name from
the annealing procedure in physical chemistry. This process starts from a metastable state of
the metal, reached by a sudden decrease of temperature; the annealing procedure consists in a
gradual cooling by which the mineral passes from the metastable phase to the low temperature
minimum of energy. During the annealing procedure the system passes  through states of thermal
equilibrium and, correspondingly, of minimal Helmholtz free energy $F$. In the limit of low
temperatures one is therefore guaranteed that the system is in the global minimum and not in a
local minimum of $F$.

A computational method trying to simulate annealing was invented about twenty years ago
\cite{SA}, based on the Metropolis algorithm \cite{metro}. This method, called {\it Simulated
Annealing} (SA) not only simulates annealing in its quest for the global minimum of the free
energy, but also in its stochastic evolution; because of this last feature it can become rather
time-consuming. This snag is avoided in Deterministic Annealing, the method we will use here.
It is still an annealing method because it points to the global minimum of $F$ and  allows
gradual cooling of the system, but it is deterministic since the procedure of optimization (see
below) is obtained deterministically and not by a random process. In the sequel we shall
describe DA in general terms, while in the next section we shall discuss the modifications we
have implemented to adapt it to the particular problem of  the determination of the four jets
in $WW$ production in $e^+e^-$ diffusion.

To start with, one defines two sets, the set of the data points $x\in X$ and the set of the
representative points $y\in Y$, also called $code-vectors$, i.e. the points that eventually
represent the clusters. One also defines a distance $d(x,y)$ between the data point $x$ and the
code-vector $y$. In jet physics $x\equiv p^{\,\mu}$, i.e. the four momentum of one of the
particles in the final state, while $y$ is the  four momentum of a cluster; as for the distance
$d(x_i,x_j)$ we will take it coincident with $y_{ij}$, i.e. one of the basic $distances$ in the
jet physics.

DA fixes the code-vectors by  minimizing the Helmholtz free energy
  \be F^*=-T\sum_x\ln\left(\sum_ye^{-d(x,y)/T} \right)\label{4} \ee
 with respect to the code vector. Basic ingredient of the calculation is the use of $F^*$ as
free energy; it is based on the use of the  conditional probabilities
 \be p(y|x)=\frac{e^{-d(x, y)/T}}{Z_x} \label{dur1}\ee with
\be Z_x=\sum_ye^{-d(x,y)/T}\ . \ee
 It is well known that the Gibbs distribution (\ref{dur1})
is the minimum point of the free energy, defined in general as \be F=D-TH\ .\label{f} \ee Here
the role of the energy  is played by \be D=\sum_{x,y}p(x,y)d(x,y)\ ,\label{8}\ee where $p(x,y)$
is the joint probability of $x$ and $y$ and the entropy is the $Shannon$ $entropy$, i.e. \be
H=-\sum_{x,y}p(x,y)\ln\,p(x,y)\ .\label{9}\ee By minimizing $F$ in (\ref{f}) one gets $F^*$
provided the Gibbs distribution is used; therefore the use of the probability (\ref{dur1}),
together with the algorithmic search for a minimum of $F^*$ in the variables $y$, is tantamount
to the quest of the global minimum of the Helmholtz free energy.

The  optimal code-vectors  $y$ are obtained by solving the equations
 \be \sum_x p(y|x)\nabla_y d(x,y)=0\ ,\label{minimum}\ee
 which is found by the above mentioned procedure of minimization of $F^*$.

During the cooling process ($T\to 0$) one encounters phase transitions that are signalled by an
increase of the number of clusters. This shows the similarity between DA and the jet clustering
algorithms, where the number of clusters  can also increase by a reduction of the parameter
$y_{cut}$.
 This aspect is not particularly relevant here, since the number of jets is held fixed.
Nevertheless we discuss it for two reasons. First, it allows to stress an important aspect of
DA that we will modify in section \ref{sec5} for the application in jet physics, i.e. the
definition of the optimal code-vectors; second, it could be useful in other applications where
the number of jets is not fixed $a$ $priori$.

To increase the number of clusters one starts with a high value of $T$; it can be shown that in
the limit of very high temperatures the minimum condition (\ref{minimum}) has one degenerate
solution (all the $y$ equal). This corresponds to one cluster and, in the statistical mechanics
analogy, to a completely disordered phase, typical of a high temperature state. After
computation of this unique value $y_1$ by (\ref{minimum}), one performs a deterministic
updating according to the formula
 \be y_1=\sum_x x p(x) p(y_1|x)\label{11}\ .\ee
 If a pre-chosen convergence test is satisfied, one decreases the temperature. One can show that
there are phase transitions when a value $y_0$ is found such that
 \be det\left[1-\frac 2 T \,C_{x|y_0}\right]=0, \label{c}\ee
 where $C_{x|y}$ is the covariance matrix computed with the (posterior) conditional probability
$p(x|y)$. If a solution of (\ref{c}) is obtained for a certain critical temperature
  \be T=T^{1\to 2}\ ,\ee
 then the previous set of code-vectors corresponds no longer to a minimum of the free energy.
Therefore one adds a new cluster (therefore there are now two code-vectors, $y_1,\,y_2$) and
the procedure of optimization is repeated.
  In general, instead of (\ref{11}) one has
\be y_j=\frac{\sum_x x p(x) p(y_j|x)}{p(y_j)}\label{11bis}\ .\ee
 The algorithm continues until a pre-fixed temperature (or number of clusters) is reached.
 It is worth noticing that this procedure does not assign uniquely each data point to a cluster,
because some points can remain unassigned \cite{DA}. Therefore, in the final step, one cools
down the system ($ T\to 0$) until all the particles are assigned.
 %As stressed already, when the number of cluster is fixed a priori
 %there is no need of these steps; they  could
 % be nevertheless implemented to control if the
 %ansatz on the number of clusters
 % does correspond to a minimum of the free energy.
\section{Deterministic annealing and W masses in
$e^+e^-$ diffusion\label{sec4}} 
\begin{figure}[ht]
\begin{center}
\includegraphics[width=9cm]{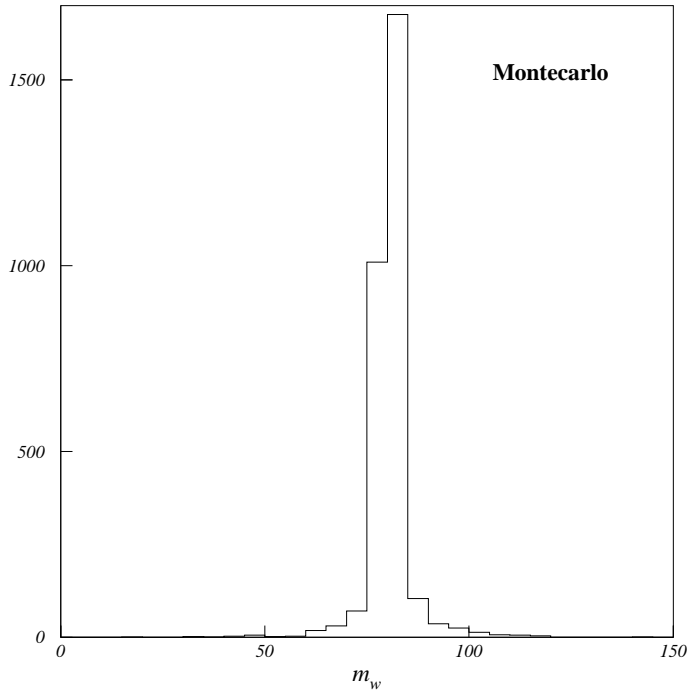}
\end{center}
\caption{The $W$-mass distribution of the actual data; $m_W$ in
GeV/c$^2$.} \label{Mass_gen}
\end{figure}

A $W$ pair created in $e^+e^-$ annihilation produces in the
subsequent evolution four jets; the study of this multi-jet final state is one of the methods
for the determination of the $W$ mass since the four jets can be divided into two pairs,
 each having as invariant mass $m^2_W$. The world average for this parameter is \cite{PDG}
\be  m_W=(80.422\pm 0.047)\ {\rm GeV/c}^2\ ,\ee
while the measurement at LEP, obtained combining both the hadronic and
semileptonic channels,
is \cite{LEP}
\be  m_W=(80.450\pm 0.039)\ {\rm GeV/c}^2\ .\ee
 Given the exploratory character of this paper we are less
 interested in the prediction of the actual experimental data than in the comparison of the DA and jet algorithms; therefore
 we choose to work with Montecarlo generated data. The data set consists
 of about 1500 events produced at the LEP energies by the KORALW
 generator \cite{mc}, which includes all the four-fermion diagrams contributing to
 $W^+W^-$-like final states. It produces the primary reference sample, with a $W$ mass of
 $m_W=80.35$ GeV (Fig.\ref{Mass_gen}). The KORALW generator is interfaced with JETSET \cite{pyt}
for fragmentation.

There is one modification to be implemented in the DA algorithm for its use in the present
study. As we have mentioned above, by eq. (\ref{11bis}) one would extract the code-vectors. Eq.
(\ref{11bis}) would produce in the present case a vector representative of the jet
 corresponding to the average 4-momentum in the jet $J$; however in all the jet clustering
 algorithms discussed in section \ref{sec2} the momentum of the single particle is compared to the
 {\it total momentum} of the jet $J$ and not to the average momentum of the particles in the
cluster $J$. In order to reproduce this feature we modify the DA algorithm by imposing only
minimization in the  probability distribution $p(y_j|x)$ and not also in the variables $y$. In
practice we use Eqns. (\ref{4}-\ref{9}), but not (\ref{minimum}) and (\ref{11bis}); in
particular we use, instead of (\ref{11bis}), \be y_j=\sum_x x p(y_j|x)\label{11ter}\ee where
$p(y_j|x)$ is still given by (\ref{dur1}).

 The practical implementation of the algorithm at fixed temperature consists of alternate updating of eqns. (\ref{11ter})
and (\ref{dur1}), until convergence is reached. The temperature is then lowered and iteration
process is restarted from the solution found at the previous temperature.
 Let us finally observe explicitly that when the temperature decreases below the critical value
$T^{3\to 4}$ the number of jets remains fixed ($K=4$). Since we go to very small temperatures,
$T\simeq 10^{-3}$, this means that the term $-T\,H$ in (\ref{f}) is negligible; therefore, the
final clusterization corresponds in practice to a minimum of the cost function
 \be D=\sum_{k=1}^4\sum_{x\in J_k}d(x,y_k)\ ,\label{8bis}\ee
where $x\in J_k$ means that the particle $x$ belong to the jet $J_k$ of code-vector $y_k$. $D$
assumes the form (\ref{8bis}) because the probabilities $p(x,y_j)$ assume only the values 1 or
0, depending on the assignment of the particle of 4-momentum $x$ to the cluster $J_k$ or to
another jet.
\section{Results and discussions\label{sec5}}
Given the data set, be it simulated or real, one has to adopt a reconstruction criterion; we
will use the following one\footnote{Other methods are discussed in \cite{paramatti}.}. Among
the three possible pairings of the four jets $J_\alpha$: $\{(J_1,J_2),(J_3,J_4)\}$,
$\{(J_1,J_3),(J_2,J_4)\}$, $\{(J_1,J_4),(J_2,J_3)\}$, with invariant masses $(m_{2k},m_{2k+1})$
($k=1,2,3$ for the three pairings), one chooses the one with the minimum value of \be
|m_{2k}-m_{2k+1}|\ .\ee This method tends to underestimate the $W$ mass, a defect that could be
corrected by different methods \cite{paramatti}; this analysis is however of no interest at the
moment.
\begin{figure}[ht]
\begin{center}
\includegraphics[width=10cm]{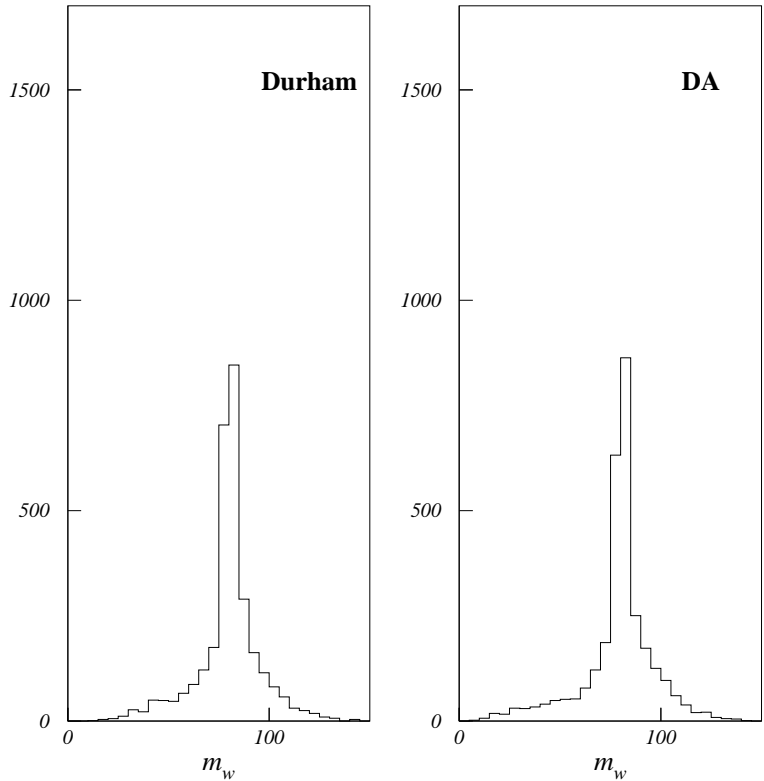}
\end{center}
\caption{The $W$-mass (in GeV/c$^2$) distributions as
reconstructed by the Durham jet algorithm (left) and Deterministic
 Annealing (right).} \label{Mass_rec}
\end{figure}
Our results are presented in fig. \ref{Mass_rec}. We plot the distribution of the $W$-mass for
the reconstructed data set obtained using the Durham  algorithm (on the left) and the
Deterministic Annealing algorithm (on the right). One can see that the two algorithms provide
similar results, giving as average value of the $W$ mass: $m_W= 79.08$ GeV/c$^2$ for the Durham
method and $m_W= 79.33$ GeV/c$^2$ for DA.

\begin{figure}[ht]
\begin{center}
\includegraphics*[width=9cm]{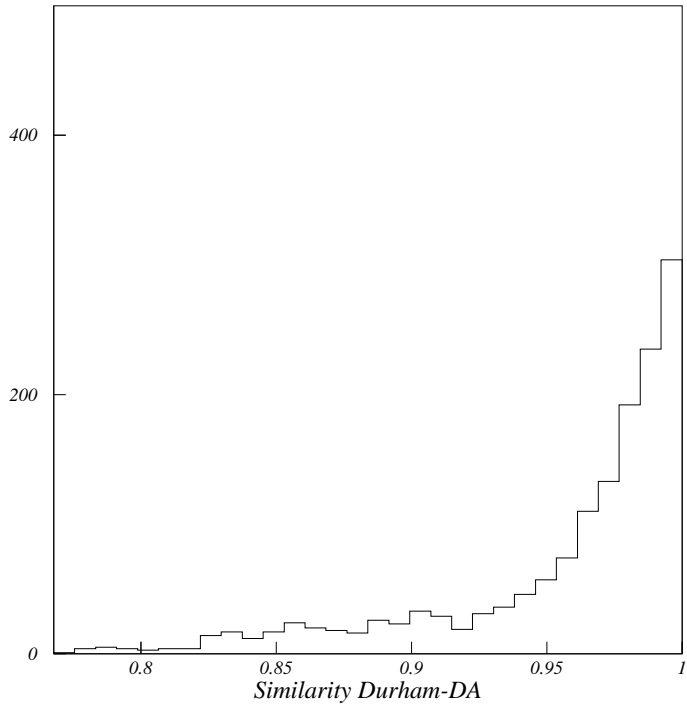}
\end{center}
\caption{The distribution of the similarity $S$.} \label{simil}
\end{figure}

The study of the differences between the two algorithms can be made using a quantitative
definition of the similarity between the two clustering procedures. We introduce the {\it
similarity parameter} $S$ by the formula:
   \be
S=\frac{\dd\sum_{i=1}^{n}\sum_{j=i+1}^{n}min\{E_i,E_j\}
\delta_{\alpha^1_{ij}\alpha^2_{ij}} }{\dd\sum_{i=1}^{n}
\sum_{j=i+1}^{n} min\{E_i,E_j\}} \label{sim_def}
   \ee
 where $\delta_{\alpha^1_{ij}\alpha^2_{ij}}$ is the Kronecker delta and
   \be \alpha^1_{ij}=1\nonumber\ee
 if the particles $i,\,j$ belong to the same cluster as defined by the Durham algorithm, while
    \be \alpha^1_{ij}=0\nonumber\ee otherwise;
$\alpha^2_{ij}$ is defined in a similar way, but with the cluster identified by DA. It is clear
that $S=1$ is equivalent to say that two particles are in the same cluster according to the
Durham method if and only if they are in the same cluster also with DA; with $S=0$ this never
happens. The factor $min\{E_i,E_j\}$ gives lower weight to pairs with at least one low-energy
particle. The histogram of the similarity $S$ is shown in Fig.\ref{simil}: we find that a
fraction of $87.8\%$ of the events have $S>0.90$, i.e. to a large extent they are clustered
identically by the two algorithms. It may be useful to observe that the percentage of events
with $S>0.80$ is $99.0\%$.
Therefore we can conclude that not only the two algorithms give similar results as far as the
average $W$ mass is concerned, as we have already observed, but also that the composition of
the jets provided by the two clustering procedures is identical in a very large fraction of the
cases.

The consequence of this analysis is twofold; on one side we are insured that in a large
majority of cases the Durham algorithm provides a minimum of the cost function (\ref{8bis});
therefore we can say that, to a large extent, the Durham algorithm is not only a local, but
also a global clusterization procedure.

On the other side, the results obtained so far show that the use of the DA algorithm might be
seen as a practical alternative to the standard jet clusterization methods in the analyses
involving huge computational tasks. This advantage may not be particularly important in the
production of $WW$ pairs at the LEP energies, but could be significant at higher energies.

The reason why we expect a significant improvement at high
multiplicities when using the DA algorithm lies in the
computational complexity of DA which is of the order $N^\alpha$
($N=$ number of particles) with $\alpha\ge 1$, in comparison with
$\alpha\ge 2$ for  the Durham algorithm. As a matter of fact in
the DA algorithm basically one performs {\it one loop} over the
particles, see e.g. eq. (\ref{4}) that contains a single loop over
$x= 1,...,N$. On the other hand, in the case of the Durham
algorithm there are {\it two loops} as shown by the discussion in
section \ref{sec2} (computation of $y_{ij}$  for $i,j=1,...,N$).
In both cases one expects values for $\alpha$ slightly larger
than, respectively, $\alpha=$ 1, 2; as a matter of fact in the DA
case there are convergence criteria at fixed temperature to be
satisfied; in the Durham case one performs other logical
operations, in particular the reordering process, which also grows
with $N$.
\begin{figure}[ht]
\begin{center}
\includegraphics*[width=10cm]{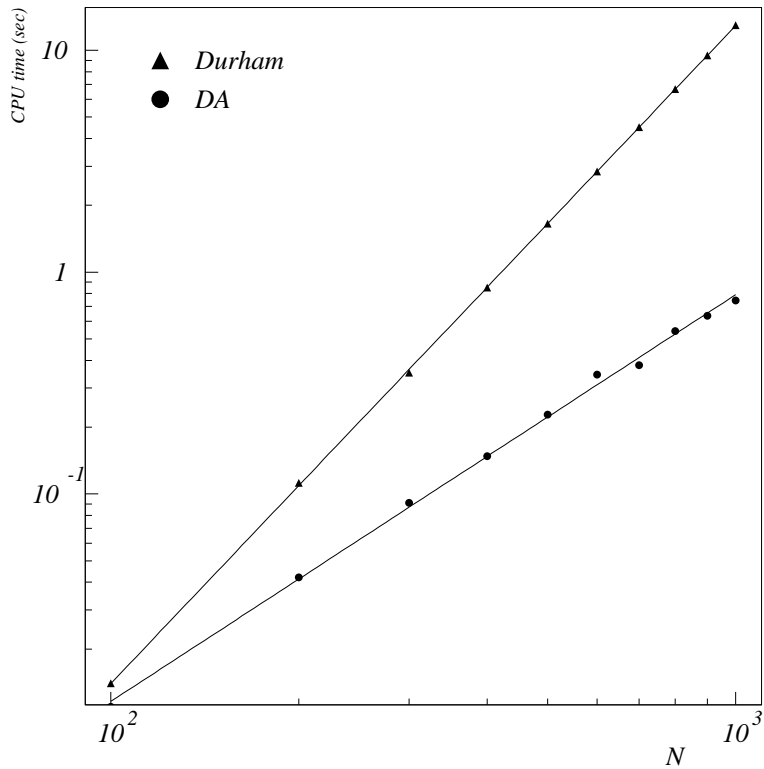}
\end{center}
\caption{CPU time (in seconds), vs. particle multiplicity for the
Deterministic Annealing algorithm (lower curve) and the Durham
algorithm (upper curve).} \label{cpu}
\end{figure}

We have performed the analysis of the computational complexity of
the two algorithms for Monte Carlo generated $e^+e^-$ events of
increasing multiplicity. The results are reported in Fig.
\ref{cpu} where we compare the CPU time (in seconds) for the DA
algorithm (lower curve) and for the Durham algorithm (upper curve)
versus the multiplicity $N$.

A fit of the two curves of fig.\ref{cpu}  gives, for large $N$,
\bea {\rm Deterministic\, Annealing}&:&~~~~~~~~~~~~~ \ t\propto
N^{1.83}\ ,\cr {\rm \,Durham\, algorithm}&:&
~~~~~~~~~~~~~~t\propto N^{2.97}\ .\eea
 These results are obtained by a processor Pentium IV 1700.

\section{Conclusions}
We have compared the results found by the standard Durham algorithm with those obtained by a
clustering algorithm based on the Deterministic Annealing procedure. The latter is a
minimization algorithm, often used in clusterization problems, and has been adapted to the
process studied here i.e. jet identification in particle production by high energy collisions.
In particular we have addressed  the study of the hadronic jets in $WW$ production by high
energy $e^+e^-$ scattering, but our results are rather general and should not depend on the
specific reaction considered. They are as follows. First, we find that the two procedures give
basically the same output as far as the clusterization is concerned. This means that one can
interpret the Durham algorithm not only as a local computational scheme, but also as a global
algorithm, i.e. a scheme attempting a minimization of a prescribed cost function. Second, we
find that the CPU time of both algorithms increases with the particle multiplicity, but the
growth is much faster for the Durham jet clustering algorithm, which is a consequence of the
higher computational complexity of the Durham scheme in comparison with Deterministic
Annealing. This result might be significant for jet physics at future accelerators such as the
Large Hadron Collider at CERN. \vskip1cm
\par\noindent{\bf Acknowledgements}
\par\noindent
We thank G. Dissertori for most useful
discussions.


\begin{thebibliography}{1-99}

\bibitem{DA}K. Rose, E. Gurewitz and G. C. Fox, Pattern Rec.
Letters {\bf 11} (1990) 589.
%
\bibitem{DA2}K. Rose, Proc. of the IEEE {\bf 86} (1998) 2211.
%
\bibitem{JADE}
W. Bartel et al. [JADE Collaboration], Z. Phys. {\bf C33} (1986)
23; S. Bethke et al. [JADE Collaboration], Phys. Lett.  {\bf B213}
(1988) 235.
%
\bibitem{Durham}
Yu. L. Dokshitzer, in  Proc. Workshop on Jet Studies at LEP and
Hera, Durham, 1990, J. Phys. {\bf G} 17 (1991) 1572ff; S. Catani,
Yu. L. Dokshitzer, M. Olsson, G. Turnock and B. R. Webber, Phys.
Lett. {\bf B269} 432; N. Brown, W.J. Stirling, Z. Phys. {\bf C53}
(1992) 629.
%
\bibitem{Cambridge}Yu. L. Dokshitzer, G. Leder, S.
Moretti and B. Webber, JHEP {\bf 9708} (1997) 001 [hep-ph/9707323].
%
\bibitem{rassegne2}S. Moretti, L. Lonnblad and T. Sjostrand,
JHEP {\bf 9808} (1998) 001 [hep-ph/9804296].
%
\bibitem{rassegne} A. Ballestrero et al., {\it Report of the QCD working group},
[hep-ph/0006259].
%
\bibitem{comparison}
M. Seymour, Z. Phys. {\bf C 62} (1994) 127;
S. Bentvelsen and I. Meyer, Eur. Phys. J. {\bf C4}(1998) 623
[hep-ph/9803322].
%
\bibitem{SA}S. Kirkpatrick, C.D. Gelatt and M.P. Vecchi,  Science {\bf
220}(1983) 671; V.\v{C}erny, J. Optimization Theory and
Applications {\bf 45} (1985) 41.
%
\bibitem{metro}N. Metropolis, A. W. Rosenbluth, M. N. Rosenbluth,
A. H. Teller and E. Teller, {\it J. Chem. Phys.} {\bf 21} (1953)
1087.
%
\bibitem{PDG} D.E. Groom et al.,
The European Physical Journal C15 (2000) 1
(http://pdg.lbl.gov/).
%
\bibitem{LEP}See the web page of the LEP Electroweak Working Group,
{\it http://lepewwg.web.cern.ch/LEPEWWG/Welcome.html} .
%
\bibitem{mc}M. Skrzypek, S. Jadach, W. Placzek and Z. Was, Comp. Phys. Commun. {\bf 94}
(1996) 216.
%
\bibitem{pyt}T. Sj\"ostrand, Comp. Phys. Commun. {\bf 82} (1994) 74;
 T. Sj\"ostrand, L. L\"onnblad and S. Mrenna, Pythia 6.2 Physics and Manual,
[hep-ph/0108264].
%
\bibitem{paramatti}
R. Paramatti, PhD thesis, Universit\`a di Roma La Sapienza, 2002.
\end{thebibliography}
\end{document}